# Theoretical and experimental study of multi-mode thermal states with subtraction of a random number of photons

Yu. I. Bogdanov[a,b,c*], N. A. Bogdanova[a,b], K. G. Katamadze[a,c,d,e],
G. V. Avosopiants[b,d,e†] and S.P. Kulik[d,e]

[a]Valiev Institute of Physics and Technology, Russian Academy of Sciences, Russia

[b]National Research University of Electronic Technology (MIET), Russia

[c]National Research Nuclear University (MEPhI), Russia

[d]Faculty of Physics, M. V. Lomonosov Moscow State University, Moscow, Russia

[e]Quantum Technology Centre, M.V. Lomonosov Moscow State University, Moscow, Russia

**ABSTRACT**

The work is devoted to the theoretical and experimental study of quantum states of light conditionally prepared by subtraction of a random number of photons from the initial multimode thermal state. A fixed number of photons is subtracted from a multimode quantum state, but only a subsystem of a lower number of modes is registered, in which the number of subtracted photons turns out to be a non-fixed random variable. It is shown that the investigation of multiphoton subtracted multimode thermal states provides a direct study of the fundamental quantum-statistical properties of bosons using a simple experimental implementation. The developed experimental setup plays a role of a specific boson lototron, which is based on the fundamental link between the statistics of boson systems and the Polya distribution. It is shown that the calculation of the photon number distribution based on the Polya's urn scheme is equivalent to a calculation using statistical weights for boson systems. A mathematical model based on the composition of the Polya distribution and thermal state is developed and verified. The experimental results are in a good agreement with the developed theory.

## 1. INTRODUCTION

From the beginning of the 20-th century to this day the thermal states of light are one of the cornerstones of quantum optics. Beginning with Planck's explanation of blackbody radiation to the present, many remarkable studies have been carried out, including Brown and Twiss measuring the angular diameter of the star Sirius[1], a ghost imaging[2–4], quantum illumination[5], thermal analog of Hong-Ou-Mandel interference[6], multiphoton superresolution[6,7], implementation of Schmidt decomposition for thermal field spatial modes[8], etc.

In recent research, conditional thermal states under photon subtraction have been used for studying various effects based on the photon correlation, such as "quantum vampire"[9,10], photonic Maxwell's "demon"[11] and allow elegant demonstrations in the field of quantum thermodynamics and information transfer[12]. They also provide a significant increase in thermal interferometry resolution[13], advantages in metrology, etc.

In our work we essentially develop our ideas related to the study of multiphoton subtracted thermal states[10,14–17]. First, we consider not only single-mode states, but also multimode ones. Second, we let the subtraction photon number to be not only a fixed number, but also a random variable with a certain probability distribution. We prepare an $M$-mode quantum state from which one subtracts a fixed number of photons $k$. Then we consider only the $m$- mode subsystem $(m \leq M)$, in which the number of subtracted photons $j$ turns out to be a random variable $(0 \leq j \leq k)$.

---

[*] bogdanov_yurii@inbox.ru
[†] avosopyantsgrant@gmail.com

The current research, in our opinion, is essential from both the fundamental and practical points of view. From a fundamental point of view, the study of multimode thermal states under photon subtraction provides a remarkable example of directly studying the fundamental quantum-statistical properties of bosons using a simple experimental setup. From a practical point of view, the considered problem provides a good testing area for quantum state engineering and debugging procedures, which is significant for quantum information technology problems.

This paper is organized as follows. In Section 2, we describe and test the developed mathematical model of the subsystem of multimode thermal field under multiphoton subtraction. In Section 3, we describe the Polya distribution, including its fundamental connection with boson system statistics. Section 4 is devoted to an experimental verification of the developed model for a single-mode subsystem case. Finally, in section 5 we summarize the main results and conclusions of this paper.

## 2. MATHEMATICAL MODEL OF THE SUBSYSTEM OF MULTIMODE THERMAL FIELD UNDER MULTIPHOTON SUBTRACTION

The main subject of this work is both theoretical and an experimental study of the statistical photon number distribution, which is described by the following generating function:

$$G(z|k,m,M,\mu_0) = (G_{th})^m F(-k,m,M,1-G_{th}), \qquad (1)$$

here $F$ is a Gauss hypergeometric function (in terms of generalized hypergeometric functions, this function is also denoted as $_2F_1$) and $G_{th} = (1+\mu_0(1-z))^{-1}$ - generating function of the single-mode thermal state.

It is assumed that we prepare the $M$-mode thermal state with a mean photon number per one mode $\mu_0$, and exactly $k$ photons are subtracted from this state. The generating function (1) corresponds to a photon number distribution of the $m$-mode subsystem ($m \leq M$).

Below we will give a detailed description of this distribution, present its statistical characteristics and describe the connection of the obtained results with the fundamental properties of boson systems.

Note, that $G(z|k,m,M) = F(-k,m,M,(1-z))$ is the generating function of the Polya distribution[18]. In our case, it describes the distribution of a random number $j$ $(0 \leq j \leq k)$ of photons registered in the $m$-mode subsystem ($m \leq M$), provided that the original $M$-mode system consists of exactly $k$ photons. A detailed description of the Polya distribution, as well as its relation to the fundamental properties of boson systems, will be presented in Section 3.

If $G_{th}$ is a generating function of single-mode thermal state, then $(G_{th})^m$ corresponds to a generating function of the $m$-mode thermal state. This directly follows from the general properties of generating functions[19] whereas the $m$-mode thermal state photon number is a sum of independent identically distributed random variables (each term is described by a generating function $G_{th}$).

The photon subtraction can be conditionally realized by transmission of the initial quantum state through a low-reflective beam splitter, combined with a single-photon detection in the reflection channel. It can be shown[15,16], that this operation leads to the following generating function transformation $G(z) \xrightarrow{\epsilon} G_1(z)$:

$$G_1(z) = \frac{G^{(1)}(z)}{G^{(1)}(1)} = \frac{G^{(1)}(z)}{\mu}. \qquad (2)$$

Here $G^{(1)}(z)$ is the first derivative of the initial generating function $G(z)$ and $\mu = G^{(1)}(1)$ is the mean photon number in the initial state. In the case of the $m$-mode thermal state $\mu = m\,\mu_0$.

Therefore, the generating function of the photon subtracted $m$-mode thermal state $G_1(z)$ can be obtained by differentiating the initial generating function $G(z) = (G_{th})^m = (1 + \mu_0(1-z))^{-m}$:

$$G_1(z) = (G_{th})^{m+1}. \qquad (3)$$

Calculating the $j$-th derivative, we obtain the generating function $G_j(z)$, corresponding to the subtraction of $j$ photons:

$$G_j(z) = (G_{th})^{m+j}. \qquad (4)$$

The equation (4) can be also used for the case of a random number of subtracted photons. This generating function, obviously, describes the sum of a random number $m+j$ terms, where each term is also a random variable with the generating function $G_{th}$.

Next, we take into account that variable $m$ (the number of registered modes) is constant, so its' generating function is $z^m$, and the variable $j$ ($0 \le j \le k$) has a Polya distribution. Then, their sum $N = m + j$ has a generating function $G_N(z) = z^m F(-k, m, M, (1-z))$. The last expression turns into the generating function (1), if the variable $z$ (the generating function of the single-photon state) is replaced with the generating function of the single-mode thermal state $G_{th}$. The correctness of such an operation is based on the generating functions composition law[19]. In our case, the composition of the generating function $G_{th}$ (random number of photons in a separate term) and the generating function $G_N(z)$ (random number of terms) leads to the calculation of a complex (compound) function $G(z) = G_N(G_{th}(z))$. The resulting complex function is identical to the desired generating function (1). Thus, the substantiation for the generating function (1) of the desired probability distribution is completed.

The considered generating function (1) leads to the following probability distribution:

$$P(n|k, m, M, \mu_0) = \frac{1}{\Gamma(m)} \frac{\Gamma(n+m)}{\Gamma(n+1)} \frac{\Gamma(M)}{\Gamma(M-m)} \frac{\Gamma(M+k-m)}{\Gamma(M+k)} \times \\ \frac{\mu_0^n}{(1+\mu_0)^{n+m}} F\left(-k, n+m, -k-M+m+1, \frac{1}{1+\mu_0}\right). \qquad (5)$$

The obtained distribution defines the probability of detecting $n$ photons in the $m$-mode subsystem, provided that exactly $k$ photons are subtracted from the complete $M$-mode thermal state.

By definition, the factorial moments of a random variable can be calculated using the generating function as follows:

$$M[k(k-1)...(k-m+1)] = \frac{\partial^m G(z)}{\partial z^m}\bigg|_{z=1}. \qquad (6)$$

The expected value is determined by the derivative of the generating function at point 1:

$$\mu = G'(1|k,m,M,\mu_0) = m\mu_0\left(1+\frac{k}{M}\right). \tag{7}$$

The second factorial moment is determined by the second derivative at point 1:

$$G^{(2)}(1) = \mu_0^2\left(m(m+1) + \frac{2km^2}{M} + \frac{2km}{M} + \frac{k(k-1)m(m+1)}{M(M+1)}\right). \tag{8}$$

The second-order autocorrelation function is:

$$g^{(2)} = \frac{G^{(2)}(1)}{\mu^2} = \frac{m(m+1) + \frac{2km^2}{M} + \frac{2km}{M} + \frac{k(k-1)m(m+1)}{M(M+1)}}{m^2\left(1+\frac{k}{M}\right)^2}. \tag{9}$$

The variance of the distribution is:

$$D = G^{(2)}(1) + \mu - \mu^2 = \mu_0^2\left(m+1 - \left(1 - \frac{mk}{M}\right)^2 + \frac{k(k-1)m(m+1)}{M(M+1)}\right) + \mu. \tag{10}$$

## 3. POLYA DISTRIBUTION AND THE FUNDAMENTAL STATISTICAL PROPERTIES OF BOSON SYSTEMS

As noted above, the Polya distribution is an important part of the distribution (1), which is the main subject of this paper. As we will see in this section, the Polya distribution reflects the fundamental collective properties of bosons. The considered properties can be studied by experimental investigation of multimode thermal states with photon subtraction using the statistical distribution (1). Thus, the problem under consideration provides a remarkable testing area for direct study of the fundamental boson quantum statistics. The fundamental connection between the statistics of bosonic systems and the Polya distribution is the basis of the experimental setup developed in the framework of this work, which plays the role of a specific boson lototron.

Historically, the Polya distribution has been found during the study of the following urn problem[20].

Let's consider an urn containing $M$ balls. Among them $m$ balls are red, and the rest $M-m$ are white. From the physical point of view it corresponds to a system of $M$ modes, within which we select an $m$-mode subsystem $(m \leq M)$. One randomly gets out a ball (subtracts a photon by means of a low reflective beam splitter). It is required to find the probability $P(j|k,m,M)$ that among $k$ removed balls exactly $j$ will be red (in the $m$-mode subsystem there will be exactly $j$ particles, provided that the initial $M$-mode system has exactly $k$ particles). There are three options described below.

a) The removed ball is returned to the urn. This system can be called a classical one, and the required probability is described by the Maxwell-Boltzmann binomial distribution with the generating function:

$$G(z|k,m,M) = \left(1 - \frac{m}{M}(1-z)\right)^k.$$

b) The removed ball is not returned to the urn. This situation corresponds to a fermion system, and it leads to the hypergeometric distribution with the generating function:

$$G(z|k,m,M) = F(-k,-m,-M,(1-z)) \tag{11}$$

c) The removed ball is returned to the urn with an additional ball of the same color. Such a scheme reflects the collective properties of the boson system. In this case the required probability is described by the Polya distribution with the generating function[18]

$$G(z|k,m,M) = F(-k,m,M,(1-z))  \qquad (12)$$

Of course, in the optical experiment the subtracted photon does not return anywhere (especially with one additional photon). From the physical point of view, this procedure means the following. The photon subtraction is a conditional procedure (it takes place, if the reflected photon is registered), and the corresponding conditional quantum states (and all their properties) may be significantly different from the initial unconditional states. These conditional states and their photon number and quadrature distributions are the main subject of this work.

Below we will focus on the latter case. In this section, we will also show that the results obtained in the considered urn scheme can be obtained by traditional methods of statistical physics based on the use of statistical weights.

Note that the parameters $m$ and $M$ in (12) do not have to be integers, which is used in practice [18].

We note also that formulas (11) and (12) for the generating functions of the hypergeometric distribution and the Polya distribution turn into one another if we change the signs of the parameters $m$ and $M$. This demonstrates the deep interconnection between these two distributions. We see that the hypergeometric distribution can be interpreted within the framework of the Polya scheme with the return and the addition of balls with a negative number of balls. Conversely, the Polya distribution can also be interpreted in the framework of the scheme without return with a negative number of balls. In our opinion, this is a certain analogy with the Dirac sea [21].

The variable $1-z$ is convenient for statistical moment calculation, and the variable $z$ – for probability calculation. The transition from the variable $1-z$ to the variable $z$ leads to the following result for the generating function of the Polya distribution:

$$G(z|k,m,M) = \frac{\Gamma(M)\Gamma(k+M-m)}{\Gamma(k+M)\Gamma(M-m)} F(-k,m,-k-M+m+1,z). \qquad (13)$$

We can consider the thermodynamic limit for a boson system, when the small subsystem of a large system is considered ($M$ and $k$ are "large", and $m$ is "small"). In the considered limit when $M \to \infty$, $k \to \infty$, and $\mu_0 = \frac{k}{M} \to const$, the Polya generating function (12) tends to the generating function of the $m$-mode boson state, which corresponds to the state of thermodynamic equilibrium:

$$F(-k,m,M,(1-z)) \to (1+\mu_0(1-z))^{-m}. \qquad (14)$$

Using the generating function, it is easy to find the probability distribution for the number of particles. In terms of the urn problem, the probability to detect exactly $j$ red balls in a sample of $k$ balls in the Polya scheme with ball return and addition equals:

$$P(j|k,m,M) = \frac{\Gamma(k+1)\Gamma(M)\Gamma(m+j)\Gamma(M-m+k-j)}{\Gamma(j+1)\Gamma(k-j+1)\Gamma(m)\Gamma(M-m)\Gamma(M+k)}. \qquad (15)$$

One can give an interpretation of this result in terms of the classical probability definition. Indeed, it is easy to see that in the case of integer values for the parameters $m$ and $M$, the distribution (15) can be written in the following equivalent form:

$$P(j|k,m,M) = \frac{C_{m+j-1}^{j} C_{M-m+k-j-1}^{k-j}}{C_{M+k-1}^{k}}. \qquad (16)$$

Here $C_{M+k-1}^{k}$ is a statistical weight in the system of $k$ bosons and $M$ levels. In standard statistical physics textbooks this result is obtained by considering the number of possible ways of placing $k$ identical balls into $M$ boxes. This number is reduced to the number of combinations in the system of $k$ points (which symbolize particles-balls) and $M-1$ partitions (which symbolize the boundaries between the levels-boxes[22]).

Statistical weight defines the denominator (16) and describes the total number of equally possible outcomes. The numerator (16) (the number of favorable outcomes) is the product of two factors: the first is the statistical weight in the system of $j$ bosons and $m$ levels in the selected subsystem, the second is the statistical weight in the system of $k-j$ remaining bosons and $M-m$ of the remaining levels.

Thus, we have two equivalent methods for calculating the probability $P(j|k,m,M)$: one based on the Polya urn scheme and the other based on the statistical weights approach.

Finally, we present the statistical moments for the Polya distribution, which can be obtained by calculating the derivatives of the generating function. The first derivative of the generating function is:

$$G'(z|k,m,M) = \frac{km}{M} F(-k+1, m+1, M+1, (1-z)). \qquad (17)$$

The expected value is determined by the generating function derivative at the point 1:

$$\mu = G'(1|k,m,M) = \frac{km}{M}. \qquad (18)$$

The second generating function derivative is:

$$G^{(2)}(z|k,m,M) = \frac{k(k-1)m(m+1)}{M(M+1)} F(-k+2, m+2, M+2, (1-z)). \qquad (19)$$

The second factorial moment is determined by the second derivative at point 1:

$$G^{(2)}(1) = \frac{k(k-1)m(m+1)}{M(M+1)}. \qquad (20)$$

The second-order autocorrelation function is:

$$g^{(2)} = \frac{G^{(2)}(1)}{\mu^2} = \frac{M}{M+1} \frac{m+1}{m} \frac{k-1}{k}. \qquad (21)$$

The variance of the distribution is:

$$D = G^{(2)}(1) + \mu - \mu^2 = \frac{km(M-m)(M+k)}{M^2(M+1)}. \qquad (22)$$

## 4. EXPERIMENTAL VERIFICATION

The quantum state with diagonal density matrix and the photon number distribution (5) for a particular case $m=1$ has been prepared and measured experimentally. The experimental setup (which is similar to the scheme presented in [16]) is shown in Fig. 1. The He-Ne cw laser beam at the wavelength of 633 nm is asymmetrically distributed by the beamsplitter BS1 between two channels: 90% of the light serves as a homodyne (local oscillator field) and 10% is used for quantum state preparation.

The initial quasi-thermal light is prepared by passing laser beam throw a rotating ground glass disk GGD[23,24]. The corresponding coherence time approximately equals the time it takes for a grain of the disk to cross the beam and can be tuned by the varying the disk rotation speed. For the presented experiment it equals $\tau_{coh} = 40\,\mu s$. Conditional photon subtraction is realized by a beam splitter with reflectivity $r = 1\%$ [25] combined with an avalanche photodiode based single-photon detector (APD) Laser Components COUNT-100C-FC with 100 Hz dark counts and a 50 ns dead time, placed in the reflection channel. Further, the prepared state comes to the homodyne detection scheme. We used the Thorlabs PDB450A balanced homodyne detector with a bandwidth of 100 kHz and a quantum efficiency of 78%. As a result, at the output of the experimental setup, we obtained the following data sets: quadrature data $P(q)$, as well as data on photocount statistics in the subtraction channel $P(N)$. Accordingly, quadrature data with photon subtraction is $P(q|N)$. All the states of light in our experiment are spatially single-mode, but the number of temporal modes can be varied.

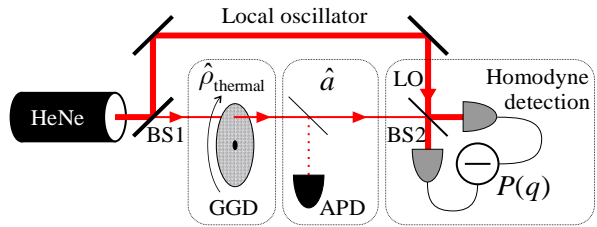

**Fig. 1.** An experimental setup for measuring quadrature distribution of one mode from multiphoton subtracted multimode thermal state. GGD is a rotating ground glass disk, APD is an avalanche photodiode based single-photon detector, BS are beam splitters.

To detect a single-mode light, we divided the time tracing into intervals (bins) of 10 µs; for each bin an integral quadrature value and the number of registered photons have been obtained. In order to avoid any interbin correlations, we selected the bins periodically separated with the period $T = 12\tau_{coh}$ [16]. Next, to perform the *k*-photon subtraction from the *M*-mode thermal state, we have combined the selected bins into the groups of *M*, and selected the groups, where the total number of subtracted photons equals *k*. All the quadrature values in all the selected groups correspond to the quadrature distribution of the single-mode (*m*=1) from the *M*-mode thermal state under *k*-photon subtraction.

Then, substituting $m = 1$ in (5), we obtain the following expression for the considered single mode photon number distribution.

$$P(n|k,M,\mu_0) = \frac{M-1}{M+k-1}\frac{\mu_0^n}{(1+\mu_0)^{n+1}} F\left(-k,1+n;2-k-M,\frac{1}{1+\mu_0}\right). \qquad (23)$$

The transition from the single-mode photon number distribution $P(n)$ to the measured quadrature distribution $P(x)$ is defined by the following expression[14–16]:

$$P(x) = \sum_{n=0}^{\infty} P(n|k,M,\mu_0)|\varphi_n(x)|^2. \qquad (24)$$

Here $\varphi_n(x)$ are harmonic oscillator eigenfunctions, defining a basic set of Chebyshev-Hermite functions.

In order to prove the adequacy of the model (24), we have prepared a set of *M*-mode thermal states under *k*-photon subtraction with $M = 1 \div 5$ and $k = 0 \div 5$. The initial mean photon number per mode $\mu_0$ for each state has been estimated with use of maximum likelihood technique [26]. Several examples of obtained quadrature distributions for $1 \div 3$-mode thermal states under 3-photon subtraction are presented in Fig. 2. The histograms correspond to the measured data, and curves – to the reconstructed states. The reconstruction has been performed on the basis of model (24) with a single free parameter $\mu_0$. The sample size $N = 10000$ for all three states. The adequacy of the experimental data has been checked using the $\chi^2$ test. One can note, that mode number increasing leads to the dramatic loss of the quadrature distribution non-Gaussianity.

The estimated values of the mean photon number $\mu = \mu_0 \left(1 + \dfrac{k}{M}\right)$ for 1, 2 and 3-mode light are presented in Fig. 3. As one can see, the required parameters reconstructed from experimental data are in good agreement with theory. We estimate the accuracy of the quantum state reconstruction using fidelity [27]:

$$F(\hat{\rho}_{des}, \hat{\rho}_{est}) = \left(Tr\left(\sqrt{\sqrt{\hat{\rho}_{des}}\hat{\rho}_{est}\sqrt{\hat{\rho}_{des}}}\right)\right)^2, \tag{25}$$

here $\hat{\rho}_{des}$ and $\hat{\rho}_{est}$ are desired and estimated density matrices, respectively. For all measured states, the fidelity was greater than 99.9%.

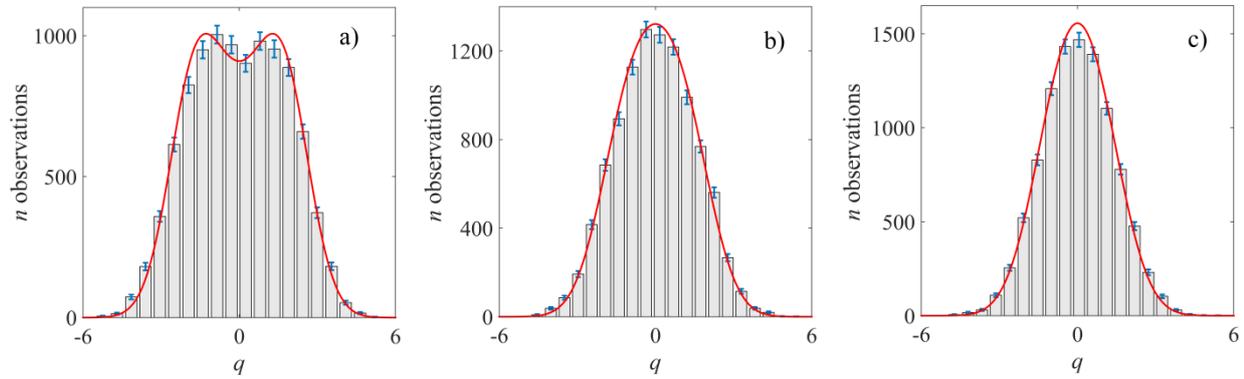

**Fig. 2.** The quadrature distributions (24) of the following states: a) $M = 1$, $k = 3$, $\mu_0 = 0.675$, b) $M = 2$, $k = 3$, $\mu_0 = 0.644$,
c) $M = 3$, $k = 3$, $\mu_0 = 0.645$. Histogram corresponds to experimental data, curve – to the model (24) with a single free parameter $\mu_0$.

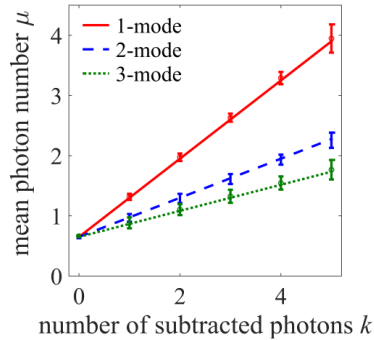

**Fig. 3.** The mean photon number $\mu$ vs. the mode number $M$ and the number of subtracted photons $k$. Solid curve corresponds to the single-mode light $(M = 1)$, dashed curve - two-mode light $(M = 2)$, dotted curve – three-mode light $(M = 3)$.

## 5. CONCLUSION

The fundamental collective properties of bosons, described by the Polya distribution can be studied through experimental research of multimode thermal states with subtraction of a random number of photons. The fundamental connection between the statistics of bosonic systems and the Polya distribution is the basis of the experimental setup developed in the framework of this work, which plays the role of a specific boson lototron. It is shown that there are two equivalent methods for calculating the photon number distribution in the boson system: one based on the Polya urn scheme and the other using statistical weights.

A mathematical model based on the Polya distribution and thermal state composition is developed and substantiated. The experimental results are in a good agreement with the theory.

The developed theory and experimental technique are essential for the quantum state engineering and debugging procedures for problems of quantum information technologies.

Acknowledgments

The work is supported by Russian Science Foundation (RSF), project no: 14-12-01338П.

Authors would like to thank Boris Bantysh for the helpful discussions.